\documentstyle[12pt]{article}

\newcommand{\sect}[1]{\setcounter{equation}{0}\section{#1}}

\def\bseq{\begin{subequation}}  
\def\eseq{\end{subequation}}
\def\bsea{\begin{subeqnarray}}  
\def\esea{\end{subeqnarray}}

\newcommand{\beq}{\begin{equation}}
\newcommand{\eeq}{\end{equation}}
\newcommand{\bea}{\begin{eqnarray}}
\newcommand{\eea}{\end{eqnarray}}
\newcommand{\non}{\nonumber}
\def\bq{\begin{quote}}
\def\eq{\end{quote}}

\newcommand{\EQ}{\begin{equation}}
\newcommand{\EN}{\end{equation}}
\newcommand{\ena}{\end{eqnarray}}

\renewcommand{\a}{\alpha}
\renewcommand{\b}{\beta}

\renewcommand{\d}{\delta}
\newcommand{\th}{\theta}

\newcommand{\pa}{\partial}
\newcommand{\di}{\partial}
\newcommand{\g}{\gamma}
\newcommand{\G}{\Gamma}

\renewcommand{\l}{\lambda}
\renewcommand{\L}{\Lambda}
\newcommand{\m}{\mu}

\newcommand{\f}{\phi}

\newcommand{\x}{\chi}

\newcommand{\s}{\sigma}
\renewcommand{\S}{\Sigma}

\renewcommand{\O}{\Omega}

\newcommand{\Db}{\bar{D}}

\newcommand{\phib}{\bar{\phi}}
\newcommand{\fb}{\bar{\phi}}
\newcommand{\gb}{\bar{\g}}
\newcommand{\Lb}{\bar{\Lambda}}
\newcommand{\Lmb}{\bar{\Lambda}}
\newcommand{\calXt}{\tilde{{\cal X}}}

\newcommand{\calX}{{\cal X}}

\newcommand{\calH}{{\cal H}}
\newcommand{\calHt}{\tilde{{\cal H}}}
\newcommand{\calHpp}{{{\cal H}}^\pp}
\newcommand{\calHtpp}{{\tilde{{\calH}}^\pp}}

\newcommand{\xpp}{\chi^\pp}
\newcommand{\xb}{\bar{\chi}}
\newcommand{\xbpp}{{\xb}^\pp}

\def\Mb{\kern 2pt\mathchoice
            {
             \vbox{\hrule width10pt height 0.4pt depth 0pt
                 \kern 1.2pt\hbox{\kern -2pt$\displaystyle M$}}}
            {
                 \vbox{\hrule width10pt height 0.4pt depth 0pt
                 \kern 1.2pt\hbox{\kern -2pt$\textstyle M$}}}
            {
\vbox{\hrule width6pt height 0.4pt depth 0pt
                 \kern 1.0pt\hbox{\kern -2pt$\scriptstyle M$}}}
            {
                 \vbox{\hrule width5pt height 0.4pt depth 0pt
                 \kern 0.8pt\hbox{\kern -2pt$\scriptscriptstyle M$}}}}

\def\Sb{\kern 2pt\mathchoice
            {
                 \vbox{\hrule width6pt height 0.4pt depth 0pt
                 \kern 1.2pt\hbox{\kern -2pt$\displaystyle S$}}}
            {
                 \vbox{\hrule width6pt height 0.4pt depth 0pt
                 \kern 1.2pt\hbox{\kern -2pt$\textstyle S$}}}
            {
                 \vbox{\hrule width3.5pt height 0.4pt depth 0pt
                 \kern 1.0pt\hbox{\kern -2pt$\scriptstyle S$}}}
            {
                 \vbox{\hrule width3pt height 0.4pt depth 0pt
                 \kern 0.8pt\hbox{\kern -2pt$\scriptscriptstyle S$}}}}

\def\Rb{\kern 2pt\mathchoice
            {
                 \vbox{\hrule width5.5pt height 0.4pt depth 0pt
                 \kern 1.2pt\hbox{\kern -2.5pt$\displaystyle R$}}}
            {
                 \vbox{\hrule width5.5pt height 0.4pt depth 0pt
                 \kern 1.2pt\hbox{\kern -2.5pt$\textstyle R$}}}
            {
                 \vbox{\hrule width3.5pt height 0.4pt depth 0pt
                 \kern 1.0pt\hbox{\kern -2.2pt$\scriptstyle R$}}}
            {
                 \vbox{\hrule width3pt height 0.4pt depth 0pt
                 \kern 0.8pt\hbox{\kern -2.2pt$\scriptscriptstyle R$}}}}

  \def\pp{{\mathchoice
          {
              \kern 1pt%
              \raise 1pt
              \vbox{\hrule width5pt height0.4pt depth0pt
                    \kern -2pt
                    \hbox{\kern 2.3pt
                          \vrule width0.4pt height6pt depth0pt
                          }
                    \kern -2pt
                    \hrule width5pt height0.4pt depth0pt}%
                    \kern 1pt
           }
            {
              \kern 1pt%
              \raise 1pt
              \vbox{\hrule width4.3pt height0.4pt depth0pt
                    \kern -1.8pt
                    \hbox{\kern 1.95pt
                          \vrule width0.4pt height5.4pt depth0pt
                          }
                    \kern -1.8pt
                    \hrule width4.3pt height0.4pt depth0pt}%
                    \kern 1pt
            }
            {
              \kern 0.5pt%
              \raise 1pt
              \vbox{\hrule width4.0pt height0.3pt depth0pt
                    \kern -1.9pt  
                    \hbox{\kern 1.85pt
                          \vrule width0.3pt height5.7pt depth0pt
                          }
                    \kern -1.9pt
                    \hrule width4.0pt height0.3pt depth0pt}%
                    \kern 0.5pt
            }
            {
              \kern 0.5pt%
              \raise 1pt
              \vbox{\hrule width3.6pt height0.3pt depth0pt
                    \kern -1.5pt
                    \hbox{\kern 1.65pt
                          \vrule width0.3pt height4.5pt depth0pt
                          }
                    \kern -1.5pt
                    \hrule width3.6pt height0.3pt depth0pt}%
                    \kern 0.5pt
            }
        }}

  \def\mm{{\mathchoice
                       {
                             \kern 1pt
               \raise 1pt    \vbox{\hrule width5pt height0.4pt depth0pt
                                  \kern 2pt
                                  \hrule width5pt height0.4pt depth0pt}
                             \kern 1pt}
                       {
                            \kern 1pt
               \raise 1pt \vbox{\hrule width4.3pt height0.4pt depth0pt
                                  \kern 1.8pt
                                  \hrule width4.3pt height0.4pt depth0pt}
                             \kern 1pt}
                       {
                            \kern 0.5pt
               \raise 1pt
                            \vbox{\hrule width4.0pt height0.3pt depth0pt
                                  \kern 1.9pt
                                  \hrule width4.0pt height0.3pt depth0pt}
                            \kern 1pt}
                       {
                           \kern 0.5pt
             \raise 1pt  \vbox{\hrule width3.6pt height0.3pt depth0pt
                                  \kern 1.5pt
                                  \hrule width3.6pt height0.3pt depth0pt}
                           \kern 0.5pt}
                       }}

\def\pd{{\kern0.5pt
                   + \kern-5.05pt \raise5.8pt\hbox{$\textstyle.$}\kern
0.5pt}}

\def\pmd{{\kern0.5pt
                  \pm \kern-5.05pt \raise6.3pt\hbox{$\textstyle.$}\kern1.5pt}}

\def\md{{\mathchoice
   {
      {{\kern 1pt - \kern-6.2pt \raise5pt\hbox{$\textstyle.$}\kern 1pt}}}
    {
      {{\kern 1pt - \kern-6.2pt \raise5pt\hbox{$\textstyle.$}\kern 1pt}}}
    {
      {\kern0.5pt - \kern-5.05pt \raise3.4pt\hbox{$\textstyle.$}\kern0.5pt}}
    {
      {\kern0.5pt - \kern-5.05pt \raise3.4pt\hbox{$\textstyle.$}\kern0.5pt}}}}

\newcommand{\Eh}{\hat{E}}

\newcommand{\Apm}{A_+^{~-}}
\newcommand{\Amp}{A_-^{~+}}
\newcommand{\Apmd}{A_{\pd}^{~\md}}

\newcommand{\Ch}{\hat{C}}

\newcommand{\Cmpp}{\Ch_{-\pd}^{~~~\pp}}

\newcommand{\Cpmm}{\Ch_{+\md}^{~~~\mm}}
\newcommand{\Cmpm}{\Ch_{-\pd}^{~~~\mm}}

\newcommand{\BAH}{\buildrel \leftarrow \over H}

\newcommand{\dirl}{\buildrel \leftrightarrow \over \di_{\pp}}

\newcommand{\baH}{\buildrel \leftarrow \over H}

\newcommand{\ad}{{\dot{\alpha}}}

\newcommand{\Del}{\nabla}
\newcommand{\Delb}{\bar{\nabla}}
\newcommand{\Delp}{\nabla_{+}}

\newcommand{\Delpd}{\nabla_{\pd}}

\def\Sc{\scriptstyle}

\newcommand{\reff}[1]{(\ref{#1})}

\newcommand{\shalf}{{\Sc\frac{1}{2}}}
\newcommand{\sihalf}{{\Sc\frac{i}{2}}}
\newcommand{\ithird}{\frac{i}{3}}
\newcommand{\sithird}{{\Sc\frac{i}{3}}}

\newcommand{\isix}{\frac{i}{6}}
\newcommand{\half}{\frac{1}{2}}
\newcommand{\ihalf}{\frac{i}{2}}
\newcommand{\stwlv}{{\Sc\frac{1}{12}}}

\renewcommand{\thefootnote}{\fnsymbol{footnote}}

\begin{document}

\newpage

\begin{titlepage}
\begin{flushright}
{hep-th/9505068}\\
{BRX-TH-370}
\end{flushright}
\vspace{2cm}
\begin{center}
{\bf {\large (2,2) SUPERGRAVITY IN THE LIGHT-CONE GAUGE}}\\
\vspace{1.5cm}
Marcus T. Grisaru\footnote{
Work partially supported by the National Science Foundation under
grant PHY-92-22318.} \\
and\\
\vspace{1mm}
Marcia E. Wehlau\footnote{\hbox to \hsize{Current address:
Mars Scientific Consulting, 28 Limeridge Dr.,
 Kingston, ON CANADA K7K~6M3}}\\
\vspace{1mm}
{\em Physics Department, Brandeis University, Waltham, MA 02254, USA}\\

\vspace{1.1cm}
{{ABSTRACT}}
\end{center}

\bq
Starting with the prepotential description of two-dimensional $(2,2)$
supergravity we use local supersymmetry transformations to go to
light-cone gauge. We discuss properties of the theory in this gauge and
derive Ward identities for correlation functions defined with respect  to
the induced supergravity action.
\eq

\vfill

\begin{flushleft}
May 1995

\end{flushleft}
\end{titlepage}

\newpage

\renewcommand{\thefootnote}{\arabic{footnote}}
\setcounter{footnote}{0}
\newpage
\pagenumbering{arabic}

\sect{Introduction}

In 1987 Polyakov \cite{Poly} examined two-dimensional induced
gravity and pointed out the existence of special features when the
theory is studied in light-cone gauge. In particular he derived a
set of Ward identities associated with the anomalous symmetries that
the theory possesses, and showed that these Ward identities
correspond  to a  ``hidden'' $SL(2,C)$ symmetry. Subsequently these
results were extended to $(1,0)$ and $(1,1)$ two-dimensional
induced supergravity \cite{MGRMX} and $(2,0)$ induced
supergravity \cite{RMX}. More recently, the Ward identities
 were used to further study properties of correlation functions
\cite{Bilal} and to determine
the dressing of $\b$-functions by induced gravity
\cite{Kogan}. In the supergravity case, similar studies were carried out
by straightforward perturbation theory \cite{MGDZ}.

For two-dimensional $(2,2)$ supergravity the exact results leading
to the Ward identities have not been available, primarily because
a formulation of the theory in terms of unconstrained superfields
had been lacking, not only in a general gauge, but even in light-cone
gauge. Recently, we have solved the constraints of $(2,2)$ supergravity
and obtained a complete description in terms of unconstrained
prepotentials \cite{MGMW} . With this solution at hand, it is possible
 to discuss the
situation in light-cone gauge, and derive appropriate
Ward identities. This is the purpose of our paper.

Unlike previously studied cases, where reaching light-cone gauge
is a fairly straightforward matter (but not quite as straightforward in the
$(0,2)$ case as the authors of ref. \cite{RMX} thought
\footnote{ A careful examination of the gauge transformations considered
by the authors of these references shows that unlike the $(1,0)$ or
$(1,1)$ cases, some components of the vielbein cannot be
straightforwardly gauged away; rather they are set to zero by the
light-cone gauge constraints.}), $(2,2)$
supergravity presents some problems, primarily because the solution
to the constraints is highly nonlinear. Furthermore, as we shall see,
whereas in other cases the unconstrained light-cone prepotential
is a general  $(1,0)$, $(1,1)$ or $(0,2)$ superfield,  here the light-cone
prepotential
 is not a full   $(2,2)$ superfield;  some of its lower components are
eliminated by a Wess-Zumino gauge choice.  As a consequence, the
nonlinearity of the theory disappears in light-cone gauge, and
 the Ward identities are as simple as in previously
studied cases.

Our paper is organized as follows: in Section 2 we review the solution of
the constraints obtained in ref. \cite{MGMW}. In Section 3 we discuss the
gauge transformations of the prepotentials. In Section 4 we use these
transformations to reach light-cone gauge, and in Section 5 we examine
the solution of the constraints in this gauge. Section 6 studies the
 light-cone gauge ``residual'' transformations  under which the induced
action changes by a local (anomalous) term. In Section 7
we use these  transformations to derive the Ward identities.
We also present some simple checks of these identities,
but the study of the ``hidden'' symmetry that these
identities represent, as well as further applications, will be postponed to a
future publication.

\sect{Solution of the (2,2) constraints}

We summarize in this section the pertinent results of ref. \cite{MGMW}.
We are considering here the $U_A(1)$ version of $(2,2)$ supergravity obtained
by setting $\S_{\a}=0$.
The spinorial covariant derivatives are
 defined by
\bea
\Del _{\a} &=& E_{\a}+ \Phi_{\a}\Lambda+\tilde{\Phi}_{\a} \tilde{\Lambda}
 \nonumber\\
&=& E_{\a} +\Omega_{\a}M +\Gamma_{\a}\Mb
\eea
with $\a=\pm$, and corresponding expressions for the complex conjugate
spinorial derivatives as well as the vectorial derivatives.
Here $ \L = M+\Mb$ and  $\tilde{\Lambda} = -i(M-\Mb )$ are the Lorentz and
$U(1)$ generators, respectively.

The vielbein is
given   by
\EQ
E_A= E_A^{~~M} \pa_M  \ .
\EN
Torsions
and curvatures are defined as usual by
\EQ
[\Del _A , \Del_B \} = T_{AB}^{~~~C} \Del_C + R_{AB}M +\Rb_{AB}\Mb
\ .
\EN
They satisfy constraints which can be described by the following
anticommutators
\bea
\{\Del_+ ,\Del_+\} &=&0 ~~~~~~~,~~~~  \{\Del_-, \Del_-\}=0 \nonumber\\
\{\Del_{+} , \Del_{\pd} \} &=&i \Del_{\pp} ~~~~,~~~~\{\Del_{-} , \Del_{\md} \}
= i \Del_{\mm} \nonumber\\
&~& \nonumber\\
  \{\Del_+, \Del_-\} &=&
-\Rb \Mb  \nonumber\\
\{\Del_+, \Del_{\md} \}&=&0
\ena
as well as their complex conjugates.

The solution of the constraints is obtained in terms of the  ``hat''
differential operators
\EQ
\hat{E}_{\pm}= e^{-H}D_{\pm}e^{H}~~~~,~~~~ H=H^mi\pa_m
\EN
with
\EQ
E_+ \equiv e^{\Sb}(\Eh _+ + \Apm \Eh_-) ~~~~,~~~~E_- \equiv  e^{\Sb}(\Eh_-+\Amp
\Eh_+)
\EN
and
\EQ
e^{\Sb} = e^{\bar{\s}} \frac{ \left[1\cdot e^{-\BAH}
 \right]^{-\frac{1}{2}}}{[1-\Apm \Amp]^{\frac{1}{2}}} E^{-\frac{1}{2}}
\EN
where $\bar{\s}$ is a covariantly antichiral superfield and $\BAH$ indicates
that the differential operator in $H^m i \pa_m$ acts on objects to its left.
 The $A$'s, as well as
the connections $\O_{\a}$, $\G_{\a}$, are given explicitly in ref. \cite{MGMW}
in terms of $H^m$ and $S$, $\bar{S}$.
The unconstrained real vector superfield $H^m$ and the chiral scalar
superfield $\s$ are the  prepotentials of $(2,2)$
supergravity.

\sect{Transformation laws for the (2,2) prepotentials}

 In obtaining the solution  of  ref. \cite{MGMW}
 we implicitly made use of the usual
 $K$ invariance of the constraints on the covariant derivatives
as well as some of the $\L$
invariance which appears as a consequence of their solution.
These invariances allow us to
eliminate the spinorial superfields $H^{\a}$,
$H^{\ad}$, as well as the imaginary part of $H^m$. We are left with further
$\L$
invariance which we discuss in the present section. The discussion follows
very much the
corresponding discussion for the 4-dimensional $N=1$ case
\cite{Superspace}. However some details are different because we avoid
the explicit introduction of  ``chiral representation''.

Covariantly chiral and antichiral scalar superfields $\Phi$, $\bar{\Phi}$ are
defined by the conditions  $\Del_{\pmd} \Phi = \Del_{\pm} \bar{\Phi}=0$ or
equivalently $\Eh_{\dot{\a}} \Phi =0$, $\Eh_{\a}\bar{ \Phi} =0$.
They may be expressed in terms of ordinary chiral and antichiral superfields
by
\EQ
\Phi = e^H \phi = e^H \phi e^{-H}~~~~,~~~\bar{\Phi} = e^{-H} \bar{\phi} =
e^{-H}\bar{\phi}e^H ~~.
\EN

The kinetic action for these superfields is
\bea
S&=& \int d^2x d^4 \th E^{-1} \bar{\Phi} \Phi = \int d^2x d^4 \th E^{-1}
\left(e^{-H} \bar{\phi}\right) \left( e^H \phi \right) \nonumber \\
&=&  \int d^2x d^4 \th E^{-1} \left(e^{-H} \bar{\phi}\right) e^{-H}\left(
e^{2H} \phi \right)
= \int d^2x d^4 \th E^{-1} e^{-H}\left( \bar{\phi} e^{2H} \phi \right)
\nonumber
\\
&=&\int d^2x d^4 \th E^{-1} e^{\BAH}\left( \bar{\phi} e^{2H} \phi \right)
\ena
where, in the last line, we have performed an integration by parts.
An equivalent expression is
\EQ
S= \int d^2x d^4 \th E^{-1} e^{-\BAH}\left( {\phi} e^{-2H} \bar{\phi} \right)
 ~~.
\EN
 Ordinary chiral and antichiral superfields transform under superspace
(coordinate) transformations as
\EQ
\phi \rightarrow  e^{i\L}\phi ~~~~,~~~ \bar{\phi} \rightarrow e^{i \bar{\L}}
\bar{\phi}
\EN
with
\bea
\L&=&\L^m i\pa_m +\L^{\a}iD_{\a} +\L^{\ad}iD_{\ad} \nonumber\\
\bar{\L}&=&\bar{\L}^m i\pa_m +\bar{\L}^{\a}iD_{\a} +\bar{\L}^{\ad}iD_{\ad} ~~.
\eea
The action in (3.2) will  be invariant provided that,
correspondingly, the prepotential $H$ transforms as
\EQ
e^{2H} \rightarrow e^{i\bar{\L}} e^{2H} e^{-i\L}
\EN
and
\EQ
E^{-1} e^{\BAH} \rightarrow  E^{-1}e^{\BAH} e^{i \buildrel \leftarrow \over
\Lb} ~~.
\EN
Equivalently
\EQ
E^{-1} e^{-\BAH} \rightarrow  E^{-1}e^{-\BAH} e^{i \buildrel \leftarrow \over
\L} ~~.
\EN

Indeed, we have then
\bea
S \rightarrow &&\int d^2x d^4 \th E^{-1}  e^{\BAH} e^{i \buildrel \leftarrow
\over \Lb}
\left( e^{i\Lb} \bar{\phi}\right) \left( e^{i\Lb}  e^{2H} \phi \right)
\nonumber\\
=&&\int d^2x d^4 \th E^{-1}  e^{\BAH} e^{i \buildrel \leftarrow \over \Lb}
 e^{i\Lb}\left(  \bar{\phi} e^{2H} \phi \right) \non \\
 =&& S
\eea
after an integration by parts of $e^{i \buildrel \leftarrow \over \Lb}$ .

In a similar manner one establishes, by examining the transformation properties
 of
a chiral  term in the action,
\EQ
S_{ch} = \int d^2x d^2 \th  e^{-2\s} \left( 1\cdot
e^{\BAH} \right) {\cal L}_{\rm chiral}
\EN
the transformation law for the covariantly chiral compensator $\s$
\EQ
e^{-2\s} \left( 1\cdot e^{\BAH} \right)\rightarrow e^{-2\s}
\left( 1\cdot e^{\BAH}
\right)
e^{i \buildrel \leftarrow \over \L} ~~.
\EN
It can be shown that indeed the constraints on the covariant derivatives
are invariant under this set of transformations.

The $\L$'s maintain the reality properties of $H$ and are restricted by two
requirements: they must
be (anti)chirality-preserving,  i.e. $D_{\pm} e^{i\Lmb} \bar{\phi}=0$
and they must maintain the vector nature of  the operator $H= H^m i\pa_m$. The
first
requirement translates into
\EQ
e^{-i\bar{\L} }D_{\pm}e^{i\bar{\L}} =a_{\pm \mp} D_{\pm} +b_{\pm \pm}D_{\mp}
\EN
or
\EQ
[\bar{\L}^M D_M, D_{\pm}]=0
\EN
summed over $M=(\pp ,\mm , \dot{+} , \md )$,
and corresponding  conditions on $\L$.  It
leads to the relations
\bea
D_{\pm}\Lmb^{\pd}&=&D_{\pm}\Lmb^{\md}=D_+\Lmb^{\mm}= D_-\Lmb^{\pp}=0
\nonumber\\
D_+\Lmb^{\pp}&=&i\Lmb^{\pd}~~~~,~~~D_-\Lmb^{\mm}=i\Lmb^{\md}
\ena
 which are solved by
\bea
\Lmb^{\pp}&=&-D_-{L}^{\pd} ~~~~,~~~~\Lb^{\pd} = iD^2{L}^{\pd}
\nonumber\\
\Lmb^{\mm}&=& -D_+{L}^{\md}~~~~,~~~~\Lmb^{\md}= -iD^2{L}^{\md} \label{L1}
\eea
in terms of an arbitrary spinor superfield ${L}^{\ad}$. Similarly, we have
\bea
\L^{\pp} &=& -D_{\md}L^+ ~~~~,~~~~\L^+= i\Db^2L^+ \nonumber\\
\L^{\mm}&=&-D_{\pd}L^- ~~~~,~~~~\L^-= -i\Db^2L^- \label{L2}~~.
\eea
The second requirement translates into the following:
for infinitesimal $\L$, $\Lmb$ we have
\bea
\d e^{2H} &=& i\Lmb e^{2H} - e^{2H}i\L \\
&=& -\Lmb^m\pa_m e^{2H} +e^{2H} \L^m\pa_m -
\Lmb^{\m}D_{\m}e^{2H} +e^{2H}\L^{\m}D_{\m}
-\Lmb^{\dot{\m}}D_{\dot{\m}}e^{2H}+e^{2H}\L^{\dot{\m}}D_{\dot{\m}}  \nonumber
\eea
To cancel the spinor derivative terms on the right hand side we require
$\L^{\m}=e^{-2H}\Lmb^{\m}e^{2H}$, $\L^{\dot{\m}}=
e^{-2H}\Lmb^{\dot{\m}}e^{2H}$, etc., i.e.
\bea
\L^{\pd}&=&ie^{-2H}D^2{L}^{\pd}e^{2H} ~~~~,~~~~\L^{\md}
=-i e^{-2H}D^2{L}^{\md}e^{2H} \nonumber\\
\Lmb^+ &=& ie^{2H}\Db^2L^+e^{-2H}~~~~,~~~~\Lmb^-=-i e^{2H}\Db^2L^-e^{-2H}~~.
\label{L3}
\eea

At the linearized level (where the covariantly chiral compensator
is equivalent to an ordinary chiral superfield)
we have for the prepotentials
\bea
\d H^{\pp}&=&\ihalf (D_{\md}L^+ -D_-{L}^{\pd} ) ~~~~,~~~~
\d H^{\mm}=\ihalf (D_{\pd}L^--D_+{L}^{\md}) \nonumber\\
\d\s &=&-\ihalf \Db^2(D_+L^+-D_-L^- ) ~~~,~~~\d
\bar{\s}=-\ihalf D^2(D_{\pd}{L}^{\pd}-D_{\md}{L}^{\md}) ~~.
\eea

\sect{Reaching light-cone gauge}

Going to a specific gauge, where certain components of gauge fields are
set to zero, involves examining their gauge transformations and showing that
for
any such transformation, $\cal{V} \rightarrow {\cal V}+\d {\cal V} = {\cal
V}+{\cal D}
{\cal L}$, one can solve for the gauge parameter ${\cal L}$ for any $\d {\cal
V}$.
We go to light-cone gauge by choosing $x^{\mm}$ as ``time'', so that
$1/\pa_{\pp}$ is
local and can be used when solving for gauge parameters without introducing
propagating
ghosts. We will show that it is possible to gauge away all of $H^{\mm}$ by
using the
gauge parameters $L^- $ and $L^{\md}$, and the
compensators $\s$, $\bar{\s}$, and certain components of $H^{\pp}$ by using
$L^+$ and $L^{\pd}$.  It is, of course,
sufficient to examine the linearized transformations.

We consider first the transformation
\EQ
\d H^\mm = \ihalf (D_\pd L^--D_+L^\md ) ~~.
\EN
We use a standard procedure to show that component by component,
components of $H^\mm$ can be gauged away by components of $L^-$ or
$L^\md$ (as usual, the vertical bar indicates evaluation at $\th^\a =
\th^{\dot{\a}} = 0$):
\bea
\d H^\mm| &=& \ihalf (D_\pd L^--D_+L^\md )| \non\\
\d D_+ H^\mm |  &=& \ihalf D_+D_\pd L^- | \non \\
\d D_-H^\mm | &=& \ihalf (D_-D_\pd L^--D_-D_+L^\md )| \non\\
&.& \nonumber\\
&.& \non\\
&.& \non \\
\d [D_+, D_\pd ] H^\mm | &=& \half \pa_\pp  (D_\pd L^-+D_+L^\md )| \non\\
&.& \non\\
&.&
\eea
Thus, for example
 (since one can invert $\pa_\pp$), the $\th^\pd$ component of $L^-$
and the $\th^+$ component of $L^\md$ can be used to gauge away the
first component and the $\th^+ \th^\pd$ components of $H^\mm$, and so
on.

It is not difficult to ascertain in this manner that $H^\mm$ can be
gauged away completely by using components of $L^-$ and $L^\md$. It
is also easy to check that once this has been achieved one cannot
use $\d \s \sim \bar{D}^2 D_- L^-$
to gauge away any component of the compensators.

We turn now to the gauge transformations induced by $L^+$ and $L^\pd$ .
Evidently one could use them to gauge away all of $H^\pp$
(at the cost of introducing propagating ghosts from the inversion of
the operator $\pa_\mm$), and this would take us to superconformal
gauge. Instead, we will use them to gauge away the compensators, and some
components of $H^\pp$.

{}From $ \d \s =- \ihalf \bar{D}^2D_+L^+$ it is easy to see that the
four independent components of $\s$, namely $\s |$, $D_+ \s |$, $D_- \s |$
and $D^2\s |$ can be gauged away. Once this has been done, we consider
the transformations $\d H^\pp = \ihalf (D_\md L^+ -D_- L^\pd )$, but with
the gauge parameters restricted now by the requirement that $\d \s = \d
\bar{\s}=0$, i.e.
\EQ
\bar{D}^2 D_+L^+ =0 ~~~~,~~~ D^2D_\pd L^\pd =0 ~~.
\EN
The first equation implies
\EQ
D_+ \bar{D}^2 D_+ L^+ =- i \pa_\pp D_+ D_\md L^+=0
\EN
and since we can remove the $\pa_\pp$ operator, $D_+
(D_\md L^+)=0$.
Together with the obvious $D_\md (D_\md L^+)=0$ statement, it follows
that  $\calX^\pp  \equiv D_\md L^+$ is a twisted chiral superfield.
Similarly, the second equation above implies that $\calXt^\pp
=D_-L^\pd$ is a twisted antichiral superfield
\footnote {We thank Martin Ro\v{c}ek and Warren Siegel for suggesting
this approach.}.
Thus, after gauging away the compensator, the residual gauge
transformations of $H^\pp$ are
\EQ
\d H^\pp = \calX^\pp +\calXt ^\pp ~~.
\EN

We proceed again by gauging away components of $H^\pp$:
\bea
\d H^\pp | &=& (\calX^\pp +\calXt ^\pp) | \non \\
\d D_+ H^\pp | &=& D_+ \calXt ^\pp| \non \\
\d D_-H^\pp | &=& D_-\calX^\pp| \non\\
\d D_\pd H^\pp | &=& D_\pd \calX^\pp | \non \\
\d D_\md H^\pp | &=& D_\md \calXt^\pp | \non \\
\d [D_+, D_\pd ] H^\pp | &=& i\pa_\pp (\calX^\pp -\calXt ^\pp) | \non \\
\d [D_-, D_\md ]H^\pp | &=& -i\pa_\mm (\calX^\pp -\calXt ^\pp) | \non \\
\d[D_+,D_\md ] H^\pp | &=& [D_+,D_\md ] \calXt ^\pp | \non \\
\d [D_-, D_\pd ] H^\pp | &=& [D_-,D_\pd ] \calX^\pp | \non \\
\d D^2 H^\pp | &=& 0 \non \\
\d \bar{D}^2 H^\pp | &=& 0 \non \\
\d \bar{D}^2 D_+ H^\pp | &=& -i \pa_\pp D_\md \calXt ^\pp | \non \\
\d \bar{D}^2D_- H^\pp | &=& i\pa_\mm D_\pd \calX^\pp |\non \\
\d D^2 D_\pd H^\pp | &=& -i\pa_\pp D_- \calX^\pp | \non \\
\d D^2 D_\md H^\pp | &=& i\pa_\mm D_+ \calXt^\pp | \non \\
\d \{ D^2 , \bar{D}^2 \} H^\pp | &=& 0 ~~.
\eea

It is clear then that we can gauge away all the components of $H^\pp$
which do not multiply  $\th^-$,  $\th^\md$, or $\th^-\th^\md$.
Of the remainder,  $D_-\calX^\pp$ appears in
the transformation of two of the components of $H^\pp$,
namely $D_- H^\pp |$ and $D^2D_\pd H^\pp |$. In fact it can be
used to go to a gauge where
$D^2 D_\pd H^\pp | -  \ihalf \pa_\pp D_- H^\pp |=0 $.
A similar statement holds with respect to $D_\md \calXt^\pp$.
It is not difficult to ascertain then that we can choose a gauge where
the prepotential has the following form:
\bea
H^{\pp} &=& \th^-\th^{\md}[h_{\mm}^{~\pp} +\th^+ \psi_{\md}^{~\pp}
-\th^{\pd}\psi_-^{~\pp}
-\th^+\th^{\pd}D^{\pp}] \nonumber\\
&&+\th^-[ e^{\sihalf \th^+\th^{\pd}\pa_{\pp}}(\l_-^{\pp}+\th^+N^{\pp})]
+\th^{\md}[e^{-\sihalf \th^+\th^{\pd}\pa_{\pp}}(\l_{\md}^{\pp}
+\th^{\pd}\bar{N}^{\pp})] \label{H1} ~~.
\eea
Thus, in light-cone gauge, the superfield $H^{\pp}$ has a natural decomposition
in terms of $(2,0)$ superfields, one of them real, the other two chiral and
antichiral, with respect to $D_\pd$ and $D_+$.
 Absorbing the $\th^-$, $\th^{\md}$ into these superfields,  we
shall write
\EQ
H^{\pp} \equiv {\cal H}^{\pp} +\chi^{\pp} +\bar{\chi}^{\pp} \label{H2}
\EN
\EQ
{\cal H}^{\pp}=\th^-\th^{\md} {\cal H}_{\mm}^{~\pp} ~~~,~~~ \chi^{\pp}=
\th^-\chi_-^{~\pp}
{}~~~,~~~ \bar{\chi}^{\pp}=\th^{\md} \chi_{\md}^{~\pp} ~~.
\EN
We note the following:
\bea
&& (\calH )^2= (\xpp )^2 = (\xbpp )^2 =0 \nonumber\\
&&\calHpp \xpp = \calHpp \xbpp =0 \nonumber\\
&& \calHpp D_- \xpp = -(D_-\calHpp)\xpp \nonumber\\
&& \calHpp D_{\md}\xbpp = -( D_\pd \calHpp)\xbpp\nonumber\\
&&D_{\dot{\pm}} \xpp = D_\pm \xbpp=0  ~~. \label{410}
\eea
Finally, we recall that we use a notation where the corresponding differential
operators
would be denoted as
\beq
\calH = i\calHpp \pa_{\pp}~~~~,~~~\chi = i \xpp \pa_{\pp}~~~~,~~~
\bar{\chi}=i\xbpp \pa_{\pp} ~~.
\eeq
Corresponding to the vanishing of the superfield products above,
 products of such operators vanish as well.

\sect{(2,2) supergravity in the light-cone gauge}

In this section we re-examine the solution to the (2,2) supergravity
constraints discussed in
ref. \cite{MGMW} and  work out the explicit form
various quantities take in light-cone gauge.
 We find it useful to split up $e^H$ as follows:
\bea
e^H &=& e^{(\calH + \x + \xb)} \non \\
&=& e^{\calH} e^{\xb} e^{\x} e^{-\shalf[\xb, \x]}
\eea
where we have used the Baker-Campbell-Hausdorff formula $e^{A+B} = e^A
e^B e^{-\shalf[A,B]}$ (all higher order terms being zero in this case).
The commutator can also be written as
\bea
 -\half[\xb, \x] &=& -\half[i \xbpp \di_\pp, i \xpp \di_\pp] \non \\
      &=& \ihalf (\xpp \dirl \xbpp)i \di_\pp
\eea
so that
\bea
e^H &=& e^{(\calHpp + \sihalf \x \dirl \xb)i \di_\pp} e^{i \xbpp \di_\pp}
           e^{i \xpp \di_\pp}  \nonumber \\
&=& e^{\calHtpp i \di_\pp} e^{i \xbpp \di_\pp} e^{i \xpp \di_\pp} \non \\
&=& e^{\calHt} e^{\xb} e^{\x}  ~~.
\eea

We have defined
\beq
\calHtpp = \calHpp + \ihalf \xpp \dirl \xbpp
\eeq
and the corresponding operator.

Note that $[\calHt, \x] = [\calHt, \xb] = 0$ because products of these
operators
vanish. Therefore we can rewrite $e^H$ as $e^{\xb}
e^{\calHt} e^{\x}$ instead, and using this and the vanishing of powers of the
superfields, we can derive simple expressions
for the $\Eh_\a$'s.  We obtain:
\bea
\Eh_+ &=& D_+ + i D_+(\calHt + \x)^\pp \di_\pp  \nonumber \\
\Eh_\pd &=& D_\pd - i D_\pd(\calHt + \xb)^\pp \di_\pp  ~~.
\eea
The remaining $\Eh$'s are somewhat more complicated. We obtain
\bea
\Eh_- &=& D_- + i D_-(\calHt + \x)^\pp \di_\pp + \xpp(\di_\pp D_- \xpp) \di_\pp
     + \calHtpp(\di_\pp D_- \xpp )\di_\pp -
(D_- \xpp)(\di_\pp \calHtpp)\di_\pp
 \nonumber \\
   \Eh_\md &=& D_\md - iD_-(\calHt + \xb)^\pp \di_\pp + \xbpp(\di_\pp D_\md
\xbpp)           \di_\pp
+\calHtpp (\di_\pp D_\md \xbpp)\di_\pp - (D_\md \xbpp)(\di_\pp \calHtpp)\di_\pp
\nonumber\\
&{}&
\eea
We note that since  $\pa_{\mm}$ does not appear in these operators,
quantities such as $\Cpmm$ and $\Cmpm$ of ref. \cite{MGMW}  vanish, and this
implies
\beq
\Apm = \Apmd =0 ~~~.
\eeq

Computing   $\{\Eh_-, \Eh_\pd\}$  we obtain $\Cmpp$, and we find
\bea
{A_-}^+ &=& i\Cmpp   \non \\
        &=& E^{-1}\left(2 D_- D_\pd \calHtpp - 2i(D_-D_\pd \di_\pp
\calHtpp)\xpp
           + 2i(D_-D_\pd \calHtpp) \di_\pp \xpp \right.\non \\
&&\left. +iD_- \xpp {\dirl} D_\pd \xbpp + \xpp(\di_\pp D_- \xpp) {\dirl} D_\pd
\xbpp \right)  ~~~.
\eea

{}From the expression in eq. (5.13) of ref. \cite{MGMW} it follows that in
light-cone gauge ${\rm sdet} \check{E}_A^{~~B}=1$ and from eqs. (5.7), (5.10),
of that reference we
work out the vielbein determinant $E$, and its inverse, which is given by
\beq
E^{-1} = 1 - [D_\pd, D_+]\calHtpp -i \di_\pp(\x -\xb)^\pp +
2 \pa_\pp \xpp \pa_\pp \xbpp +i D_+ \xpp \dirl D_\pd \xbpp ~~.
\eeq

We also work out the explicit expressions
\bea
1\cdot e^{-{\baH}} &=& 1 - i \di_\pp \calHtpp-i\di_\pp \xpp-i \di_\pp \xbpp -
 \di_\pp \xpp \di_\pp \xbpp
     - \xpp{\di}^2 _\pp \xbpp  \non\\
(1\cdot e^{-{\baH}})^{-1} &=& 1 + i \di_\pp
\calHtpp +i\di_\pp \xpp +i \di_\pp \xbpp -
 \di_\pp \xpp \di_\pp \xbpp
     + \xpp {\di}^2_\pp \xbpp  ~~.
\eea
{}From (2.7) it follows then that
\beq
e^{2\Sb}
=1 + 2D_+ D_\pd \calHtpp+ 2i \di_\pp \xbpp + \xpp {\di}_\pp^2 \xbpp + \di_\pp
      \xpp \di_\pp \xbpp + i D_+ \xpp {\dirl} D_\pd \xbpp
\eeq
and
\beq
e^{2S} = 1 - 2 D_\pd D_+ \calHtpp - 2i \di_\pp \xpp + \xbpp {\di}_\pp^2 \xpp +
\di_\pp
            \xpp \di_\pp \xbpp + i D_+ \xpp {\dirl} D_\pd \xbpp ~.
\eeq

Therefore, from the general expressions for the connections given in
ref. \cite{MGMW}
$\O_+ = \O_{\pd}=0$, while from
\beq
\O_- = - e^{\bar{S}} \Eh_+ \Amp ~~~,~~~\G_+=2e^{\bar{S}}\Eh_+S ~~~,~~~\G_-=
-e^{\bar{S}}\Eh_-S
\eeq
  we obtain, after some algebra,
\bea
\Rb &=& -E_+ \G_- -E_-\G_+-\frac{1}{2} \O_-\G_+ \nonumber\\
&=& -4 e^{2\bar{S}}[ (\Eh_-\bar{S})(\Eh_+S)-\Eh_+\Eh_-S] \non \\
&=& -4 \Eh_- \Eh_+\left[ e^{2\bar{S}}S \right] ~~.
\eea
This expression, and a corresponding one for the conjugate,
 can be rewritten in the pleasing form
\beq
\bar{R}	= 4 \Del^2 S ~~~~,~~~ R= 4\Delb ^2 \bar{S}~~~.
\eeq
In terms of our explicit light-cone prepotential we find then that
\bea
\Rb &=& -2 e^{-i(\x + \calHt) \di} D^2 [ 2i \di_\pp
 (\calHt + \x)^\pp - 4 \di_\pp\xpp\di_\pp\xbpp
            - \xpp {\di}_\pp^2 \xbpp - \xbpp {\di}_\pp^2 \xpp]  \non \\
R &=& -2 e^{i(\xb + \calHt)\di} {\Db}^2 [-2i \di_\pp (\calHt + \xb)^\pp
 - 4 \di_\pp\xpp\di_\pp\xbpp
            - \xpp {\di}_\pp^2 \xbpp - \xbpp {\di}_\pp^2 \xpp] ~.\non \\
&&~
\eea

More compactly, in terms of the original superfield $H^{\pp} = \calHpp +\xpp
+\xbpp$
this can be rewritten as
\beq
R= 4i e^H \bar{D}^2 \pa_{\pp} e^{-H} H^{\pp} ~~~~,~~~
\Rb= -4i e^{-H}D^2\pa_{\pp}e^{H} H^{\pp} ~~.
\eeq

\sect{Light-cone gauge transformations}

In the next section  we shall derive Ward identities for correlation functions
defined by (functional) averaging with the (nonlocal)  induced supergravity
action. They are obtained by making use of the invariance of the
functional integral under a change of variables which is a field
transformation.
The only requirement is that under this transformation the variation of the
induced action should be local. This will be the case if the field
transformation is
a gauge transformation for which the induced action is anomalous.  In the
present context this will be true for the general gauge transformations of the
prepotentials, restricted however to light-cone gauge,
and chosen  to preserve
the form of $H^{\pp}$ in \reff{H1}. We begin by deriving
these transformations. We note that, unlike the considerations of the previous
section, which only required knowledge of the linearized transformations,
here we need the full nonlinear transformations. Because of the nature of the
light-cone prepotentials, this is not a difficult task.

We consider the general gauge transformation
\EQ
e^{2H} \rightarrow e^{i \Lb} e^{2H} e^{-i\L}
\EN
with the general form of the parameters $\L$, $\Lb$ given in
\reff{L1}-\reff{L3}
{}.
We choose however $L^- = L^{\md}=0$, and we must  suitably
restrict the form of $L^+$ and $L^{\pd}$ so as to preserve the form of
$H^{\pp}$ in
\reff{H1}. To see how this works we look first at the linearized level.
We must restrict $\d H^\pp = \ihalf (D_\md L^+-D_-L^\pd )$
so that
\EQ
\d H^\pp | = D_+H^\pp |=D_\pd H^\pp | = [D_+,D_\pd] H^\pp |=0
\EN
and require as well that the gauge transformations preserve the
(anti)chirality with respect to $D_\pd$, $D_+$ of the
 terms proportional to $\th^-$, and $\th^\md$, i.e. of $\xpp_-$
and $\xpp_\md$, respectively.
It follows that
\bea
D_-{L}^\pd &=& \bar{\a}^\pp +\th^{\md}  \bar{\eta}_{\md}^\pp-
\ihalf \th^- \th^\md \pa_{\mm} \bar{\a}^\pp  \nonumber\\
D_\md L^+ &=& {\a} ^\pp +\th^- {\eta}_-^\pp +\ihalf \th^-
\th^\md \pa_\mm {\a}^\pp \label{DL}
\eea
or \bea
L^+= \th^{\md}  ({\a}^\pp + \th^- {\eta}_-^\pp ) \nonumber\\
L^\pd = \th^-(\bar{\a}^\pp  +\th^{\md}{\eta}_{\md}^\pp)
\eea
with
${\a}(\th^+, \th^{\pd}) =\bar{\a}(\th^+, \th^{\pd})$,
and  $\eta_-^\pp (\th^+ , \th^{\pd})$ chiral. (However, this last
condition will be modified at the nonlinear level).
We denote $\th^- \eta_-^\pp \equiv \eta^\pp$.

To obtain the full light-cone gauge transformations we start again with the
Baker-Campbell-Hausdorff formula, but keeping higher
order terms  in the prepotentials. (However, since we only need infinitesimal
transformations, we can work to first order in the gauge parameters, while
terms of order higher than second in the prepotential vanish). We apply
\beq
e^A e^B = e^{A+B+ \shalf[A,B] + \stwlv[[A,B],B] + \stwlv[[B,A],A]}
\eeq
twice to the expression $e^{i\Lb} e^{2H} e^{-i\L}$.
We obtain
\beq
e^{2H} \rightarrow e^{i(\Lb - \L) + 2H + i[\Lb + \L, H] + \sithird [[\Lb - \L,
     H], H]} ~~.
\eeq
Therefore,
\beq
\d H = \ihalf (\Lb-\L) + \ihalf [\Lb+\L, H] + \isix [[\Lb-\L, H],H] ~~.
\eeq
Substituting in the expressions for $\L, \Lb$ and $H$, we obtain
\bea
\d H^{\pp}\pa_{\pp} &=& \ihalf (\Lb-\L)^\pp \di_\pp + \ihalf
(\Lb-\L)^+ D_+
+\ihalf (\Lb-\L)^\pd D_\pd -\half (\Lb+\L)^M(D_MH^\pp)\di_\pp \non \\
&&+ \half H^\pp[\di_\pp(\Lb+\L)^\pp]\di_\pp + \half H^\pp[\di_\pp(\Lb+\L)^+]
D_+  + \half H^\pp[\di_\pp(\Lb+\L)^\pd]D_\pd  \non \\
&& - \frac{i}{6} \{(\Lb-\L)^M(D_M H^\pp)(\di_\pp H^\pp)\di_\pp -H^\pp \di_\pp
[(\Lb-\L)^M D_M H^\pp]\di_\pp  \non \\
&& - H^\pp[\di_\pp(\Lb-\L)^\pp](\di_\pp H^\pp)\di_\pp +H^\pp\di_\pp[H^\pp
\di_\pp(\Lb-\L)^\pp]\di_\pp  \non \\
&& - H^\pp[\di_\pp(\Lb-\L)^+](D_+ H^\pp)\di_\pp +H^\pp\di_\pp[H^\pp\di_\pp
(\Lb-\L)^+]D_+    \non \\
&& - H^\pp[\di_\pp(\Lb-\L)^\pd](D_\pd H^\pp)\di_\pp +H^\pp\di_\pp[H^\pp\di_\pp
(\Lb-\L)^\pd] D_\pd \} ~~.
\eea
Since the left hand side only contains vector derivatives we require the
spinorial derivative terms on the right hand side to vanish.
Collecting  the coefficients of $D_+$ and
$D_\pd$, respectively, we obtain the following two equations:
\bea
(\Lb - \L)^+ - iH^\pp \di_\pp(\Lb + \L)^+ - \frac{1}{3}
 H^\pp \di_\pp \left[H^\pp \di_\pp (\Lb-\L)^+ \right] &=& 0  \\
(\Lb - \L)^\pd - iH^\pp \di_\pp(\Lb + \L)^\pd -
\frac{1}{3}H^\pp \di_\pp \left[ H^\pp  \di_\pp (\Lb-\L)^\pd
\right] &=& 0 ~~.
\eea
These equations express, in light-cone gauge, the  conditions given in
\reff{L3}.

Similarly, looking at the vectorial pieces (the coefficients of $\di_\pp$), we
find:
\bea
\d H^\pp &=& \ihalf (\Lb-\L)^\pp - \half (\Lb+\L)^\pp(\di_\pp H^\pp) -\half
(\Lb+\L)^+
      (D_+ H^\pp)  \non \\
&& -\half (\Lb+\L)^\pd(D_\pd H^\pp) + \half H^\pp
 \di_\pp(\Lb+\L)^\pp  \non \\
&&-\isix ( \Lb - \L )^+ D_+ H^\pp \pa_\pp H^\pp
  -\isix ( \Lb - \L )^\pd D_\pd H^\pp \pa_\pp H^\pp \non \\
&&+\ithird H^\pp \pa_\pp ( \Lb - \L )^+ D_+ H^\pp
  +\ithird H^\pp \pa_\pp ( \Lb - \L )^\pd D_\pd H^\pp \non \\
&&+\isix H^\pp ( \Lb - \L )^+ \pa_\pp D_+ H^\pp
  +\isix H^\pp ( \Lb - \L )^\pd \pa_\pp D_\pd H^\pp \non \\
&& - \isix (\Lb-\L)^\pp(\di_\pp H^\pp)(\di_\pp H^\pp)
 + \isix H^\pp\di_\pp(\Lb-\L)^\pp
     ( \di_\pp H^\pp) \non \\
&& + \isix H^\pp(\Lb-\L)^\pp  \di_\pp^2 H^\pp - \isix H^\pp H^\pp \di_\pp^2
    (\Lb-\L)^\pp ~~.
\eea
However, many of the terms in these expressions may be dropped because
of \reff{410}.

We need expressions for $(\Lb \pm \L)^M$. Using the restricted forms of
$L^+$, $L^\pd$ in \reff{DL},
we have
\bea
(\Lb - \L)^\pp &=& [{\a} - \bar{{\a}} +\th^- \eta_- - \th^\md
{\bar{\eta}}_\md +
          \ihalf \th^- \th^\md \di_\mm ({\a} + \bar{{\a}})  ]^\pp \\
(\Lb + \L)^\pp &=& -[{\a} + \bar{{\a}} + \th^- \eta_- + \th^\md
{\bar{\eta}}_\md
+ \ihalf \th^- \th^\md \di_\mm ({\a} - \bar{{\a}})]^\pp
\non  \\
(\Lb - \L)^+ &=& 2i [iH^\pp\di_\pp - \di_\pp (\xpp\xbpp)\di_\pp -
 2\xpp\xbpp
\di_\pp^2]
  D_\pd({\a} + \th^- \eta_- + \ihalf \th^- \th^\md \di_\mm {\a})^\pp \non  \\
(\Lb + \L)^+ &=& 2i[1+iH^\pp\di_\pp - \di_\pp (\xpp\xbpp)\di_\pp
 -2 \xpp\xbpp
\di_\pp^2]
  D_\pd({\a} + \th^- \eta_- + \ihalf \th^- \th^\md \di_\mm {\a})^\pp \non
\non\\
(\Lb - \L)^\pd &=& -2i[-iH^\pp\di_\pp - \di_\pp (\xpp\xbpp)\di_\pp
 -2 \xpp\xbpp
\di_\pp^2]
  D_+(\bar{{\a}} + \th^\md {\bar{\eta}}_\md - \ihalf \th^- \th^\md \di_\mm
\bar{{\a}})^\pp
   \non \\
(\Lb + \L)^\pd &=& 2i[1-iH^\pp\di_\pp - \di_\pp (\xpp\xbpp)\di_\pp
 - 2\xpp\xbpp
\di_\pp^2]
  D_+(\bar{{\a}} + \th^\md {\bar{\eta}}_\md - \ihalf \th^- \th^\md \di_\mm
\bar{{\a}})^\pp  \non
\eea
We find then
\bea
\d H^\pp &=& \ihalf \left({\a}- \bar{\a} +
\eta - \bar{\eta} +\ihalf \th^- \th^\md \pa_\mm ({\a}
 + \bar{\a}) \right)^\pp \nonumber\\
&&+\frac{1}{2}\left( {\a}+ \bar{\a} +\eta  +\bar{\eta}
 +\ihalf \th^- \th^\md \pa_\mm ({\a} - \bar{\a})
\right)^\pp \pa_{\pp}H^\pp \nonumber\\
&&-i \left( [1+iH^\pp \pa_{\pp}]D_\pd ({\a}+\eta )^\pp
\right) D_+H^\pp \nonumber\\
&& -i\left( [1-iH^\pp \pa_{\pp}]D_+ (\bar{\a}+\bar{\eta} )^\pp \right)
 D_\pd H^\pp \nonumber\\
&&-\frac{1}{2} H^\pp \pa_{\pp}( {\a} +\bar{\a} +\eta
+\bar{\eta})^\pp ~~. \label{13}
\eea
Since the left hand side does not contain terms independent of  $\th^-$,
$\th^\md$, or
 $\th^- \th^\md$, we require
\beq
{\a}^\pp - \bar{ \a}^\pp=0 ~~.
\eeq
We have used this fact, as well as \reff{410},
 in dropping some terms on the RHS of \reff{13}.

We consider separately the variation of the
terms proportional to $\th^-$, to  $\th^\md$, and to $\th^- \th^\md$.
We have
\bea
\d \xpp &=& \ihalf \eta^\pp + {\a}^\pp \pa_{\pp} \xpp
-i D_\pd {\a}^\pp D_+ \xpp - \xpp
\pa_{\pp} {\a}^\pp
\nonumber\\
\d \xbpp &=& -\ihalf \bar{\eta}^\pp + {\a}^\pp \pa_{\pp} \xbpp
 -iD_+ {\a}^\pp D_\pd \xbpp -\xbpp \pa_{\pp} {\a}^\pp ~~.
\eea
Since the superfields $\xpp$, $\xbpp$ are chiral and antichiral
respectively,
we obtain restrictions on the parameters from
$D_\pd \d \xpp = D_+ \d \xbpp = 0$,  which turn out to be
\bea
&&D_\pd (\half \eta^\pp +i \xpp \pa_\pp {\a}^\pp)= 0 \nonumber\\
&&D_+(\half \bar{\eta}^\pp -i\xbpp \pa_\pp {\a}^\pp)=0
\eea
and are solved by
\bea
\eta^\pp &=& -2i \xpp \pa_\pp {\a}^\pp+ D_\pd D_+ (\xpp \a^\pp )
+2D_\pd \g_\md^\pp \nonumber\\
 \bar{\eta}^\pp&=& 2i \xbpp \pa_{\pp} {\a}^\pp  -D_+ D_\pd (\xbpp \a^\pp )
+2 D_+ \bar{\g}_-^\pp \label{eta}
\eea
with arbitrary $\g$, $\bar{\g}$.
Then
\bea
\d \xpp &=& iD_{\pd}[\g_\md^\pp  + \half \xpp D_+ \a^\pp
- \half  {\a}^\pp D_+ \xpp ] \nonumber\\
\d \xbpp &=& -i D_+ [ \bar{\g}_-^\pp +\half {\a}^\pp D_\pd \xbpp
-\half \xbpp D_\pd \a^\pp  ]~~. \label{chi}
\eea

Turning to the variation of the $\th^- \th^\md$ terms, a certain amount of
algebra
leads to the conclusion that the quantity which transforms simply is not
$\calHpp$ but
\beq
 \check{\calH}^{\pp} = \calHpp +i \xpp \dirl \xbpp ~~.
\eeq
We find
\beq
\d \check{\calH}^{\pp} = -\half \th^- \th^\md \pa_\mm {\a}^\pp +
{\a}^\pp \pa_{\pp}\check{\calH}^{\pp}
 -iD_\pd {\a}^\pp
D_+ \check{\calH}^{\pp} -iD_+{\a}^\pp D_\pd \check{\calH}^\pp
 -\check{\calH}^{\pp}  \pa_{\pp}{\a}^\pp ~. \label{dH}
\eeq
Eqs. \reff{chi} and \reff{dH} represent the final form of the residual
light-cone
transformations.

\sect{Light-cone gauge Ward identities}

The induced $(2,2)$ supergravity action has the form
\EQ
S_{ind}= \frac{c}{4\pi} \int d^6z \bar{R} \frac{1}{\Box_c}R
\EN
where $\Box_c$ is a suitable d'Alembertian defined, e.g., by obtaining the
induced
action from integrating out the covariantly chiral  scalar (Goldstone)
superfield in
\EQ
S_G= -\frac{c}{4\pi} \int d^6z \bar{\Psi}\Psi +\int d^4z \Psi R + \int
d^4\bar{z} \bar{\Psi} \bar{R}
\EN
(i.e. $\Box_c$ is the inverse of the operator $\Delb^2 \Del^2$ acting on a
covariantly chiral superfield).

We  consider correlation functions of the supergravity fields, and also of
matter fields,
in the presence of the induced action, defined by
\EQ
<X(z_1,z_2,...z_n)>= \int {\cal D} (H, \phi )e^{S_{ind}(H) +S_m(\phi ) }
X(z_1,z_2,...z_n)
\EN
where $X(z_1,z_2,...z_n)$ stands for a product of $n$ supergravity or matter
fields.
We work in light-cone gauge and in the functional integral make a change of
integration variables which is a $\L$-transformation. We assume that under this
transformation
$S_m$ is invariant, while the induced action varies into the (local) anomaly.
We
obtain the Ward identity
\bea
0&=&  \int {\cal D} (H, \phi  )e^{S_{ind}(H) +S_m(\phi ) } \left[  \d  S_{ind}
+\sum _i \d _i X(z_1,z_2,...z_n) \right] \nonumber\\
&=& < \d S_{ind} X(z_1,z_2,...z_n) > + \sum_i < \d_i X(z_1,z_2,...z_n)>
\eea
where $\d_i X(z_1,z_2,...z_n)$ is the variation of the $i$'th field in
$X(z_1,z_2,...z_n)$.

In the light-cone gauge the functional integration is over the chiral and
antichiral superfields $\xpp$ and $\xbpp$ and the real superfield $\calHpp$,
and the gauge transformations are
given in \reff{chi} and \reff{dH}. For the matter
 fields, e.g. for chiral scalar
superfields with weight $\l$  we assume transformations of the form
$\d \phi = i[\L, \phi ] +i \l(1.{\buildrel \leftarrow \over \L} )\phi$ which
reduce in the
present situation to $\d \phi =  -i\bar{D}^2 (L^+D_+ \phi )
+i \l (\bar{D}^2 D_+ L^+)\phi$.

In general the induced action varies into
\bea
\d S_{ind} &=&\frac{ic}{4\pi}\int d^2x d^4 \th ~E^{-1}[ R\Del_+L^+
+\bar{R} \Delb_{\pd}{L}^{\pd}
] \nonumber\\
&=& \frac{ic}{\pi}\int d^2x d^4 \th E^{-1} [\Del_{\pd}\Del_+S \Del_-L^{\pd}
+ \Del_+ \Del_{\pd} \bar{S} \Del_{\md}L^+ ] ~~.
\eea

We compute this variation explicitly, making use of the specific restricted
forms of the various superfields. After some straightforward algebra
we find
\bea
\Delpd \Delp S&=& -i D_\pd D_+ \pa_\pp \calHpp
+\pa^2_\pp \xpp -D_+\xpp \pa_\pp^2 D_\pd \xbpp
+D_+\pa_\pp^2 \xpp D_\pd \xbpp \nonumber\\
&& -2 D_+ \pa_\pp \xpp D_\pd \pa_\pp \xbpp
+i\pa_\pp ( \pa_\pp \xbpp \pa_\pp \xpp )
\eea
and
\bea
\Del_{+}\Del_\pd \bar{S}&=&
i D_+ D_\pd \pa_\pp \calHpp -\pa^2_\pp \xbpp  -D_+\xpp \pa_\pp^2 D_\pd \xbpp
+D_+\pa_\pp^2 \xpp D_\pd \xbpp \nonumber\\
&& +2 D_+ \pa_\pp \xpp D_\pd \pa_\pp \xbpp
+i\pa_\pp (\pa _\pp \xpp \pa_\pp \xbpp ) ~~.
\eea
Since these expressions explicitly contain one factor of $\th^-$ and
$\th^{\md}$
respectively, we can simplify
\EQ
E^{-1} \sim 1-i\pa_{\pp}(\xpp - \xbpp ) ~~.
\EN

We choose as our transformation parameters the superfields
\bea
L^+ &=& \th^{\md} ({\a}^\pp + \th^- \eta_-^\pp ) \nonumber\\
L^\pd &=& \th^-(\bar{\a}^\pp  +\th^{\md}\bar{\eta}_{\md}^\pp)
\eea
with  ${\a}^\pp  (\th^+, \th^\pd ) = \bar{{\a}} ^\pp (\th^+, \th^\pd )$
and $\eta_- ^\pp(\th^+, \th^\pd )$,
$\bar{\eta}_{\md}^\pp (\th^+, \th^\pd )$ satisfying \reff{eta}, consistent
with the restriction on the
transformation parameters that we have discussed earlier. Because of the
explicit $\th^-$, $\th^\md$ dependence there are a number of simplifications,
e.g. $\Del_\md L^+ \sim D_\md L^+$, etc.. After  substitution in the variation
of the induced action, and some algebra, we obtain
\bea
\d S_{ind}&=& \frac{ic}{\pi} \int d^2x d^4 \th
 \left\{\left(i[D_+,D_\pd] \di_\pp \check{\calH}^\pp + \di_\pp^2
 (\xpp -\xbpp) \right)\a^\pp \right. \non \\
&&\left. ~~~~~~~~~~+2 \pa_\pp^2 \xpp D_+\bar{\g}_-^\pp
 - 2\pa_\pp^2 \xbpp D_\pd \g_\md^\pp \right\} ~~. \label{dS}
\eea
We note that in this expression the term $ \di_\pp^2 (\xpp
- \xbpp)\a^\pp $ can be dropped because it is missing either a $\th^-$ or a
$\th^\md$ and therefore gives zero upon $d^4 \th $ integration.

We have now all the ingredients for writing down the Ward identities.

We consider, as an explicit example which can be verified in perturbation
theory,
the correlator  $<\phib (y) \phi(z)>$ for an ordinary
chiral scalar superfield ($\l =0$). The matter lagrangian is
\EQ
S_m  =\int d^6z E^{-1}
\left(e^{-H}\bar{\phi}\right) \left(  e^H \phi \right) = \int d^6z
( \bar{\phi} \phi -2 H^\pp D_\pd \bar{\phi} D_+ \phi +\cdots ) ~~.
\EN

The Ward identity becomes:
\beq
<\d S_{ind} \fb \f> + <\d \fb \phi> + <\phi \d \fb> = 0
\eeq
with $\d S_{ind}$ given by \reff{dS}, and $\d \phi$ and $\d \fb$ given by
\bea
\d \phi &=& -i{\Db}^2 [\th^\md (\a^\pp - 2i \xpp \di_\pp \a^\pp
+D_\pd D_+  (\xpp \a^\pp )+ 2 D_\pd
           \g_\md^\pp) D_+ \phi] \non\\
\d \fb &=&-i D^2 [\th^-(\a^\pp + 2i \xbpp \di_\pp \a^\pp
-D_+D_\pd  (\xbpp \a^\pp )+ 2 D_+ \gb_-^\pp)
           D_\pd \fb] ~~.
\eea
  Simple counting (order by order in $c$) reveals that terms linear in $c$
trivially satisfy the Ward identity, terms independent of $c$ lead to
tree graphs, and terms proportional to $1/c^L$ give rise to loops.
We consider here the tree graphs and note that terms dependent
on $\a$ and $\g$ must separately satisfy the Ward identity.  We obtain the
 following equation for the $\a$-dependent part :
\bea
\lefteqn{\frac{ic}{\pi} \int d^2x d^4 \th <i[D_+, D_\pd] \di_\pp \calH^\pp(x)
 \a^\pp(x) \fb(y) \f(z)>}  \non \\
&&-i <D^2 [\th^- \a^\pp D_\pd \fb(y)] \f(z)> -i <\fb(y) \Db^2 [\th^\md \a^\pp
D_+ \f(z)]> = 0
\eea
where $< ......>$ indicates expectation value with respect to $S_{ind} +
S_m$.

In the first term we substitute the interaction term from $S_m$ to first
order in $\calHpp$ and compute
\bea
\lefteqn{-\frac{2ic}{\pi} \int d^2 x d^4 \th_x d^2 w d^4 \th_w ~\a^\pp (x)
{}~\cdot} \non\\
&& < \left (i [D_+,D_\pd]\pa_\pp
\calHpp (x) \right) \calHpp (w) D_\pd \phib (w) D_+ \phi (w) \phib (y)
\phi (z) >_0
\eea
Wick-contracting the $\calH^\pp$'s and the $\phi$'s. The second and third
terms are evaluated to zeroth order in the interaction.

We need the corresponding propagators. For the matter
fields we have the standard chiral propagators, e.g.
\EQ
< \phib (w) \phi (z) > = - \frac{D^2 \Db^2}{\pa_\mm \pa_\pp} \d^{(2)}
(w-z) \d^{(4)}(\th_w -\th_z ) ~~.
\EN
To obtain the $\calH^\pp$ propagator we return to the induced action which,
to quadratic order, reduces to
\EQ
S_{ind}^{(2)} = -\frac{2c}{\pi} \int d^2 x d^4 \th
\left( \Db^2 \calHpp \frac{\pa_\pp}
{\pa_\mm} D^2 \calHpp + \xbpp \pa_\pp^2 \xpp \right) ~~. \label{S2}
\EN
We consider the first term only,   split off the $\th^-$, $\th^\md$
factors in $\calHpp = \th^- \th^\md \calHpp_\mm (\th^+ \th^\pd )$,
 and perform the corresponding
integration to obtain
\EQ
S_{ind}^{(2)} = -\frac{2c}{\pi} \int d^2x d \th^+ d\th^\pd D_\pd \calHpp_\mm
D_+ \calHpp_\mm = -\frac{c}{\pi} \int d^2x d \th^+ d\th^\pd
\calHpp_\mm [D_+,D_\pd ] \calHpp_\mm ~~.
\EN
in (2,0) superspace.  Using
$[D_+,D_\pd ]^2 = -\pa_\pp^2$
we obtain for the propagator
\EQ
< \calHpp_\mm (x) \calHpp_\mm (w) > = \frac{\pi}{2c}
\frac{\pa_\mm}{\pa_\pp^3} [D_+ , D_\pd ] \d^{(2)}(x-w)
\d ( \th^+_x -\th^+_w)  \d ( \th^\pd _x -\th^ \pd _w) ~~.
\EN

In the first term of the Ward identity we also separate off the
corresponding $\th$'s, but instead of integrating them out we write
\EQ
\th^-_x \th^\md_x \th^-_w \th^\md_w =
\th^-_x \th^\md_x \d (\th^-_x -\th^-_w ) \d ( \th^\md_x -\th^\md_w )
\EN
The last two $\d$-functions can be combined with those in the
$\calHpp_\mm$ propagator leading effectively to
\EQ
< \calHpp (x) \calHpp (w) > = \frac{\pi}{2c} \th^-_x \th^\md_x
\frac{\pa_\mm}{\pa_\pp^3} [D_+ , D_\pd ] \d^{(2)}(x-w)\d^{(4)}(\th_x-\th_w) ~~.
\EN
Now, ordinary $D$-algebra can be carried out
in standard fashion, verifying the Ward identity which is
depicted graphically in Fig.1.
\vspace{5mm}

\let\picnaturalsize=N
\def\picsize{6.0in}
\def\picfilename{lc1.eps}
\ifx\nopictures Y\else{\ifx\epsfloaded Y\else\input epsf \fi
\let\epsfloaded=Y
\centerline{\ifx\picnaturalsize N\epsfxsize \picsize\fi
\epsfbox{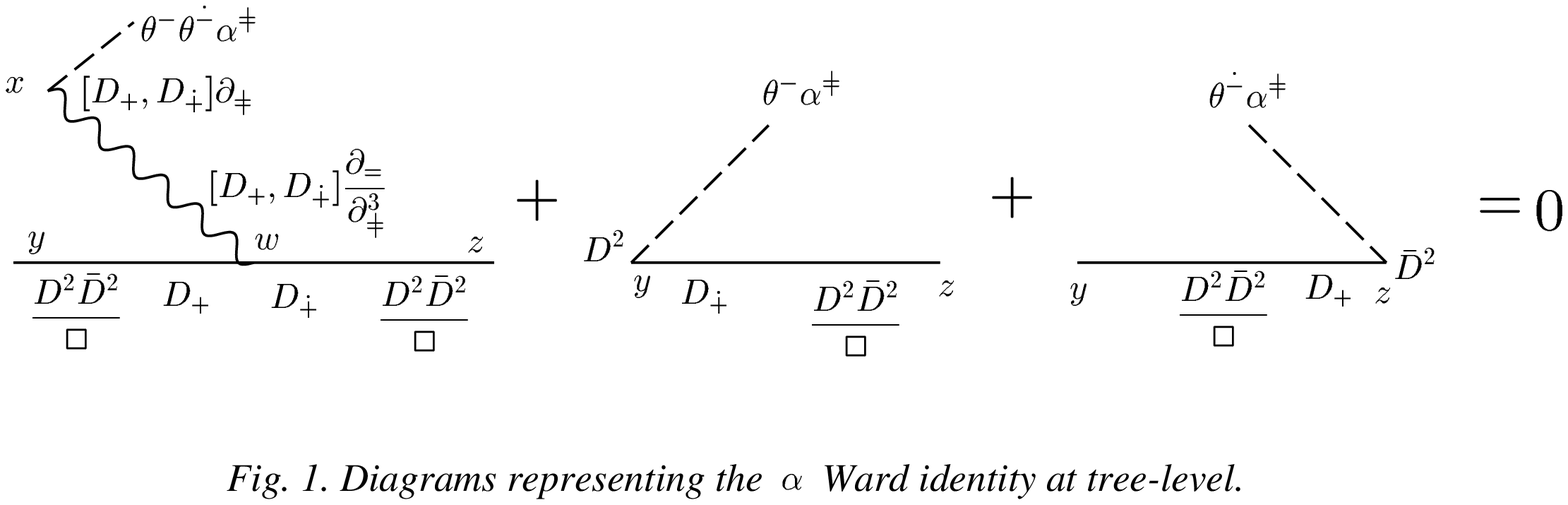}}}\fi

A similar treatment of the terms proportional to $\gb$ can be used
to verify the corresponding part of the Ward identity
\bea
\frac{2ic}{\pi} \int d^2x d^4 \th < \di_\pp^2 \xpp (x)  D_+ \gb_-^\pp (x)
  \fb(y) \f(z)>
-2i <D^2 [\th^- D_+ \gb_-^\pp  D_\pd \fb(y)] \f(z)> =0\non\\
\eea
To first order  in the interaction  we obtain for the first term
\bea
\lefteqn{ -\frac{4ic}{\pi} \int d^2x d^4 \th_x d^2 w d^4 \th_w  ~\cdot} \non\\
&& < \di_\pp^2 \xpp (x)  D_+ \gb_-^\pp (x)
\xbpp (w) D_\pd \fb (w) D_+ \f (w)  \fb(y) \f(z)>_0   ~~.
\eea

We separate out from the $\x$ and $\xb$ fields factors of  $\th^-$ and
$\th^\md$
and obtain the $\x$-$\xb$ propagator in $(2,0)$ superspace from \reff{S2}:
\EQ
< \xpp_- (x) \xbpp_\md (w) > =- \frac{\pi i}{2c}
\frac{D_\pd D_+}{\pa_\pp^3}\d^{(2)}(x-w)
\d ( \th^+_x -\th^+_w)  \d ( \th^\pd _x -\th^ \pd _w) ~~.
\EN
In the Ward identity we take out of the $\x$'s corresponding factors of $\th$
as well as a $\th^\md_x$ from $\gb$
and write  $\th^-_x \th^\md_x \th^\md_w= -\th^-_x\th^\md_x
\d  (\th^\md_x-\th^\md_w) D_- \d (\th^-_x- \th^-_w) $. Combining these
$\d$-functions
with the ones in the propagator we obtain the diagrammatic representation
 in Fig. 2,  and the Ward identity can be verified by straightforward
$D$-algebra.
\vspace{5mm}
\let\picnaturalsize=N
\def\picsize{4.3in}
\def\picfilename{lc2.eps}
\ifx\nopictures Y\else{\ifx\epsfloaded Y\else\input epsf \fi
\let\epsfloaded=Y
\centerline{\ifx\picnaturalsize N\epsfxsize \picsize\fi
\epsfbox{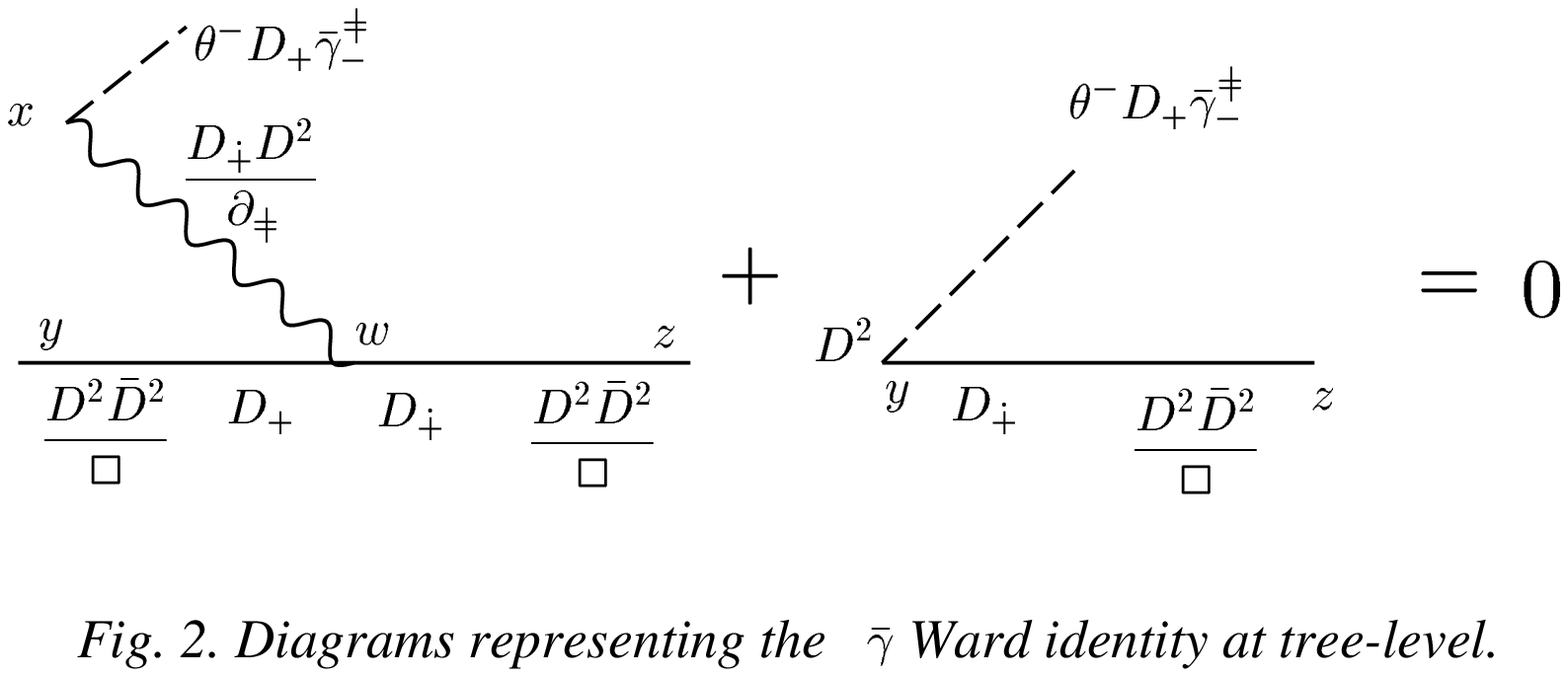}}}\fi

\sect{Conclusions}
In this work, starting with the prepotential formulation of $(2,2)$
supergravity, we have used the gauge transformations of the
theory to go to light-cone gauge. In this gauge the dependence
on the prepotentials of the
various geometrical quantities is (almost) linear and this allows us
to exhibit them explicitly.

The main interest in the light-cone formulation rests on the possibility
of investigating properties of the induced (due to the superconformal
anomaly) nonlocal action $S_{ind}$, and in particular discovering the
underlying ``hidden'' symmetry of the resulting theory,
 generalizing to the $(2,2)$ case the $SL(2C)$ symmetry
discovered by Polyakov for the induced gravity case \cite{Poly}.

We have obtained the transformation laws which are
relevant to the light-cone gauge, and written down the general
form of the corresponding anomalous Ward identities. However,
the explicit solution of these identities, the study of the
corresponding underlying algebra, and applications such as
the study of $\b$-function dressing \cite{Kogan,MGDZ} are
left to a future publication.

{\bf Acknowledgments} We thank Martin Ro\v{c}ek and Warren Siegel
for discussions and  Daniela Zanon for participation in the early stages of
this work. M.E.W. thanks the Physics Department of Queen's University
for hospitality.

\end{document}